# Extended Power Method to Calculate
# Pre-Selectable Eigenvalues and Eigenstates

**Dr. Wolfgang Berger**

Taunusblick 13
D 61479 Glashütten
Tel. 49(0)6174-63305
Fax 49(0)6174-62922
Email wb@wberger.com

*Abstract*

*The quantum mechanical expression relating two commuting operators is reformulated such that the power method (also called method of moments) for iteratively calculating eigenvalues and eigenvectors becomes applicable. The new iterative scheme thus obtained allows to pre-select a quantum number of one of the two commuting operators and then calculate a corresponding eigenvalue of the other operator. The result is the common eigenvector of the eigenvalue pair. Among others the method may be used to calculate excited states, cope with degeneracy and/or accelerate convergence. Small example calculations are presented as a "Proof-of-Concept" and to reveal some properties of the new method.*

## 1    Introduction

Calculating eigenvectors and eigenvalues of hermitian matrices or operators has been a major effort since the beginning of quantum mechanics [1]. Iterative methods became the focus of "eigen"-calculations because they are

- Conceptually easy,
- Applicable in matrix representations, particle hole formulations, coordinate space or abstract operator representations in a Hilbert space [2] and have the
- Potential to combine optimization of parameters with eigenpair-iterations [3]

Among the well known methods in this field are the power method [4] (also called method of moments) and it's extensions like 2x2-iterations [5] (which is equivalent to steepest descent), the conjugate gradient method or the Lanczos tridiagonalisation [1,4].

Common to the various approaches is that they essentially aim at dominant states - the ground state (and at most a few low lying states) or some high lying states. Here "dominant" refers to the magnitude of the eigenvalue. Simultaneous eigenstates of two or more commuting operators are taken into account by selecting proper start vectors and/or applying projection techniques.

The present note deals with an extension of the power method which generates iteratively simultaneous eigenvalue pairs and the corresponding eigenstate of two commuting operators. The desired quantum state of one of the operators may be pre-specified. The new scheme can be derived from basic quantum mechanical principles and reformulated into the form of a generalized eigenvalue problem [1] which in turn suggests to apply the well known power method.

## 2    Power Method and Extensions

Let H be a hermitian operator. (To be specific, we may think of H as being a Hamiltonian.) Then the power method to calculate the dominant eigenvalue e (the largest in magnitude) of H is defined by the following recipe:

$$e_n|n+1> = H|n>, \quad <n|n> = <n+1|n+1> = 1, \quad e_n = ||H|n>|| \rightarrow e, \quad n = 1,2,3 ... \quad (1)$$

To calculate the smallest eigenvalue (in magnitude) one may apply *inverse iteration* which means to apply the same iteration scheme but using the inverse of H instead of H:

$$1/e_n|n+1> = H^{-1}|n> \iff H|n+1> = e_n|n>, \quad 1/e_n = ||H^{-1}|n>||, \quad n = 1,2,3 ... \quad (2)$$

Although it may seem at a first glance that knowledge of $H^{-1}$ is necessary, the equivalent formulation in equation (2) shows, that one may alternatively solve a system of inhomogeneous linear equations in each iteration step.

Let S be a hermitian operator that commutes with H having an eigenvalue s and let $|y_e>$ and $|y_s>$ be the corresponding normalized eigenstates of H and S.

$$H|y_e> = e|y_e>, \quad <y_e|y_e> = 1 \tag{3a}$$
$$S|y_s> = s|y_s>, \quad <y_s|y_s> = 1 \tag{3b}$$
$$[H,S] = 0 \tag{3c}$$

Assuming s different from zero we use (3b) to formally define an iteration scheme for s (as done in eq. (1) for e):

$$s|m+1> = S|m>, \quad <m|m> = <m+1|m+1> = 1, \quad |m> \rightarrow |y_s>, \quad m = 1,2,3 ... \tag{4a}$$

Introducing a spectral shift µ into eq. (3b) and applying inverse iteration we get:

$$1/(s-µ)|m+1> = (S-µE)^{-1}|m> \tag{4b}$$

in analogy to eq. (2).

Note that (1) and (4a) define different tasks. In (1) the eigenvalue $e_n \rightarrow e$ is to be calculated together with its corresponding eigenvector, say $|y_e>$. Equation (4a), however, may be used as a basis to calculate an eigenvector $|y_s>$ with s being a given eigenvalue of the operator S. The modification (4b) will – as we shall see – be used to direct and modulate convergence to desired eigenvalues by properly choosing s and µ.

Assuming that under appropriate conditions both sequences converge, that is $|n> \rightarrow |y_e>$ and $|m> \rightarrow |y_s>$ the two vectors $|y_e>$ and $|y_s>$ depend on the spectrum of their respective operators H and S and the start vector. They are otherwise not related a priori. Thus the question arises whether and how one may couple both iterations such, that we can find the *simultaneous* eigenvector $|y_{e,s}>$ of H and S belonging to the eigenvalue pair (e,s) with s given.

## 3    Commuting Operators

To answer this we first introduce a spectral shift µ (µ not an eigenvalue of S) into (3c):

$$[(S-µE),H)] = 0 \tag{5}$$

Letting this operate on a *simultaneous* eigenvector $|y>$ of S with eigenvalue s and H with eigenvalue e we get (E being the unity operator)

$$e(S-µE)|y> = (s-µ)H|y>, \quad S|y> = s|y>, \quad H|y> = e|y> \tag{6a}$$

or equivalently

$$H|y> = e/(s-µ) (S-µE)|y>, \quad S|y> = s|y>, \quad H|y> = e|y>. \tag{6b}$$

We remark that if equation (6a) holds we have automatically

$$S|y\rangle = s|y\rangle \iff H|y\rangle = e|y\rangle \quad (6c)$$

which is, what we are aiming at.

Eq. (6b) has a formal structure like "$B|y\rangle = \lambda C|y\rangle$" which is known as "generalized eigenvalue problem" and which has iterative solutions under quite general assumptions [1]. Instead of going into theoretical detail, however, in the present note we proceed to derive and demonstrate an iterative scheme based on equation (6b).

## 4    Extended Algorithm

To extract an iterative method to solve (6b) we rewrite (6b) to take on the form of an eigenvalue equation for e:

$$(s-\mu)(S-\mu E)^{-1}H|y\rangle = e|y\rangle \quad (6d)$$

Formulating a power method to solve equation (6d) we get

$$(s-\mu)(S-\mu E)^{-1}H|n\rangle = e_n|n+1\rangle, \quad \langle n+1|n+1\rangle = \langle n|n\rangle = 1, \; n=1,2,3\ldots \quad (7a)$$

or equivalently

$$(s-\mu)e_n^{-1}H|n\rangle = (S-\mu E)|n+1\rangle, \quad \langle n+1|n+1\rangle = \langle n|n\rangle = 1, \; n=1,2,3\ldots \quad (7b)$$

Here s is an eigenvalue of S and μ a spectral shift which may be used to "modulate" the iterations. $e_n$ is chosen such, that the iterated vectors are normalized.

Note that if we put S = E, s = 1 and μ = 0 we recover the original power scheme (1). Setting H = E one gets inverse iteration with spectral shift for S (eq. (4b)).

As indicated earlier, the genuine power method for H yields the dominant eigenpair and if it is turned into inverse iteration it again yields the dominant eigenpair where the eigenvalue now equals the inverse of the smallest eigenvalue of H. To calculate eigenpairs not being the lowest or highest respectively, power shift plus inverse iteration may be applied to calculate an eigenvalue e of H "close" to some guessed value. Introducing a second operator S with known eigenvalues it is possible to apply one of the iterative schemes defined by (7a) or (7b) and therefore avoid the necessity of guesses of e.

The dominance property remains essentially true even if one were to iterate with a modified H like products of H and another operator: A dominant eigenvalue will very often still prevail. With the help of a second operator S and suitable choices of s and spectral shift μ we can "tune" the iteration to "neutralize" the power of dominant eigenvalues and create a new "dominant" eigenvalue to converge to.

In general it may be expected that the combination of the two algorithms – power method plus inverse iteration – bears the chance to accelerate convergence with respect to the sole application of the power method or one of its classical extensions – again with the possibility of a "fine tuning" through µ.

In the case of degenerate eigenvalues of H, the introduction of a suitable operator S may be used, to iteratively construct eigenvectors that span the corresponding subspaces in a controlled manner. Further there are potential uses like coping with "pseudo convergence" [6] (when iterated eigenvalues remain almost steady for a number of iterations suggesting convergence where there is none) or exploit the new method with an eye on projection techniques.

The price one has to pay when using equation (7a) is that one has to know $(S-µ)^{-1}$ which is in general not the case. Alternatively one may therefore use equation (7b) which can be regarded as an inhomogeneous system of linear equations for $(S-µ)$ and which may be solved by standard methods – either finite numerical procedures (like Gauss's elimination method [1]) or by applying an iterative method (like Jacobi or Gauss-Seidel [1]).

## 5  Numerical Verification

### 5.1  Minimal Example

| Input | | | |
|---|---|---|---|
| **Basis vector \|i>** | \|1> | \|2> | \|3> |
| Eigenvalues $e_i$ of \|i> | 1 | 2 | 3 |
| Eigenvalues $s_i$ of \|i> | 1.5 | 1 | .5 |
| Coordinates of Start vector \|n=0> | 1 | 1 | 1 |
| **1. After 2 iterations with s (preselected eigenvalue of S) =1 , µ = 0.9** | | | |
| $e_{n=2}$ = <n=2\|H\|n=2> | - | 2.019 | - |
| $s_{n=2}$ = <n=2\|S\|n=2> | - | 0.990 | - |
| Coefficents of \|n=2> | 0.007 | 0.990 | 0.139 |
| **2. After 2 iterations with s (preselected eigenvalue of S) =1.5 , µ = 1.4** | | | |
| $e_{n=2}$ = <n=2\|H\|n=2> | 1.081 | - | - |
| $s_{n=2}$ = <n=2\|S\|n=2> | 1.459 | - | - |
| Coefficents of \|n=2> | 0.965 | 0.241 | 0.107 |

**Tab. 1:** First example to verify the expected behavior of the new algorithm. H and S are taken as diagonal matrices containing the specified eigenvalues as diagonal elements.

As a first "proof of concept" Tab. 1 above shows results of two calculations in a minimal 3-dimensional space – just enough to exhibit the expected behavior.

In this example the usual power method would converge to the dominant vector |3> of H with eigenvalue e = 3. Specifying s = 1 as the desired eigenvalue of S and setting the shift µ close to that value, iterations deliver the pre-selected result namely convergence to vector |2> with eigenvalues $s_n \rightarrow$ s = 1 of S and $e_n \rightarrow$ 2 of H. Finally by specifying s = 1.5 as the desired eigenvalue of S and setting the shift µ close to this value, iterations again show convergence towards the pre-selected values namely $s_n \rightarrow$ 1.5 of S and $e_n \rightarrow$ 1 of H and as eigenvector we find vector |1>.

These plus some additional trials not shown here indicate, that convergence in principle as well as convergence speed strongly depend on the spectral shift – a behavior inherent in the method of inverse iterations. Further, one iteration step of the new method amounts to about two or more iteration steps of the power method. Therefore one has to be careful in comparing convergence speeds.

### 5.2 Semi-realistic Example

As a richer example a calculation in a 15-dimensional vector space is shown below. Although the real potential of iterative methods are operations in high-dimensional spaces, this size suffices to get a first impression of the general behavior of the method.

| Eigenvalues "s" of S and "e" of H in 15x15 Example ||||||||||
|---|---|---|---|---|---|---|---|---|---|
| basis | \|1> | ... | \|5> | \|6> | \|7> | \|8> | ... | \|11> | ... | \|15> |
| s | 0.3 | - | 1.5 | 1.8 | 1.9 | 2.0 | - | 2.9 | - | 4.1 |
| e | 3.4 | - | 1.0 | -0.1 | -0.2 | -2.0 | - | -3.8 | - | -5.4 |

**Tab. 2:** Eigenvalues no. 1 … 15 of 15-dimensional example calculation. H and S are taken as diagonal matrices containing the specified eigenvalues as diagonal elements.

With the system defined by Tab. 2 several calculations have been made. Here we show results for eigenvector |11> as an example of a low lying excited state and eigenvectors |6> as an example of close eigenvalues. Results are shown in Tab. 3 below.

Calculations no. (1) and no. (2) look for an eigenstate with s = 2.9 corresponding to an eigenvalue e = -3.8 of H (see Tab.2, vector |11>). Iterating with a spectral shift µ = 2.8 a good approximation is already achieved after 2 iterations with small improvements after 2 more iterations). In the last column of Tab. 3 the coefficient of eigenvector |11> in the iterated vector is shown. For the exact solution it is equal to 1.

| Some Results of 15x15 Example | | | | | | |
|---|---|---|---|---|---|---|
| Calc.-no. | Eigen-vector no. | Pre-selection s | Spectral Shift μ | Iter.-no. n | $s_n$ | $e_n$ | $c_n$ |
| (1) | \|11> | 2.9 | 2.8 | 2 | 2.895 | -3.7871 | 0.98363 |
| (2) | \|11> | 2.9 | 2.8 | 4 | 2,8998 | -3.7996 | 0.99968 |
| (3) | \|6> | 1.8 | 1.78 | 9 | 2.0 | -2.0 | 0.00461 |
| (4) | \|6> | 1.8 | 1.79 | 9 | 1.85 | -0.657 | 0.82814 |
| (5) | \|6> | 1.8 | 1.795 | 4 | 1.8 | -0.10661 | 0.99839 |

**Tab. 3:** Typical Convergence Patterns of 15-dimensional sample. The start vector components are all ones.

Whereas the previous two examples represent a "standard" situation, calculations no. (3) to no. (5) are (as expected) more delicate. As can be seen, the critical point is the selection of he spectral shift. In the case of μ = 1.79 there will be proper convergence, but at a very low speed. After a small change to μ = 1.795 we are back to rapid convergence again. This demonstrates the possibility to enhance convergence speed by adjusting the spectral shift.

In the case of μ = 1.78 convergence to the wanted eigenvector isn't achieved at all: Iterations converge to state |8>. This can be explained by decomposing the iterated states in terms of simultaneous eigenstates |i> of H and S with coefficients $c_i$ and eigenvalues $\varepsilon_i$ and $\sigma_i$. Inserting this into eq. (7a) we see, that in each iteration the new coefficients contain a factor $\varepsilon_i/(\sigma_i - \mu)$. Convergence is directed to the eigenstate |i> where the factor is maximal.

## 5.3 Convergence Patterns

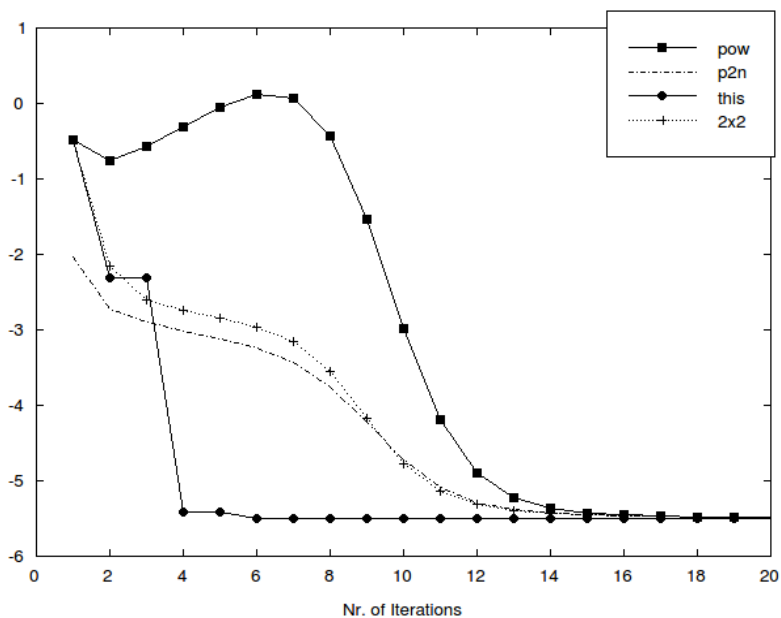

**Fig. 1:** Pseudo Convergence

Fig. 1 shows the results of a calculation with intentionally chosen extreme starting conditions to provoke a phenomenon we call "pseudo convergence" [6]. One can obtain such a behavior through setting certain coefficients in the start vector to extremely high/ low values. In our example we have chosen high eigenvalue-components as dominant.

Curve "pow" shows the expectation values <n|H|n> (<n|n> = 1) as approximations to the eigenvalues corresponding to the approximate eigenvectors |n> in iteration step n of the plain power method. The curve even raises before finally converging to the exact ground state.

The corresponding "2x2" curve performs of course better than "pow" and shows, why we call this convergence pattern "pseudo convergence": At a certain point it looks as if the curve were to converge while it is actually only at a small "rest" at an intermediate level.

Although not directly relevant in the present context we have included a plot of ||H|n>|| ("p2n"). It can be seen, that this number lies substantially below the power method and even below the 2x2-method. The reason is, that this quantity is of order $H^{2n}$. Since H|n> is available in both methods at actually no additional cost, it is advisable to keep the "p2n"-values in mind in situations where it is not clear whether we encounter pseudo convergence or not.

The convergence pattern of "this", the algorithm considered in this paper, is dramatically fast and seems to almost ignore the possibility of pseudo convergence. This happens however at the cost of additional information to be provided at the beginning i.e. the wanted eigenvalue of operator S and a suitable shift. (Note that in the diagram points at even iteration numbers represent the result after the inverse operation step and those at odd iteration numbers the results after the power -method step.)

**Fig. 2:** Convergence to an Excited State

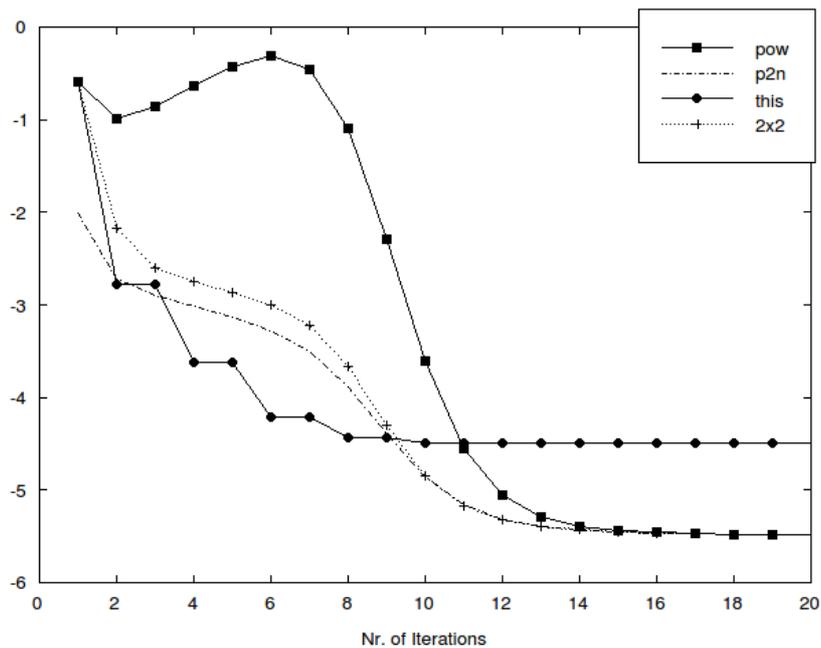

Fig. 2 finally shows a situation (curve "this") where convergence of the algorithm is directed to an excited state by specifying an appropriate eigenvalue s of S and a suitable spectral shift µ. For comparison, the ground state iterations of the previous diagram are included.

## 5      Summary and Discussion

This note suggests an extension to the power method for calculating eigenvalues and eigenvectors of hermitian operators. The new method is derived from the fundamental quantum mechanical principle of commuting operators representing compatible observables. The quantum theoretical formulation of this principle is then transformed into a form known as "generalized eigenvalue problem". Finally one of the methods to solve this problem - the power method - is applied to the quantum mechanical formulation. In this way a new iterative algorithm has been defined.

The primary task of the new method is to calculate an eigenvalue e and the corresponding eigenvector of an operator H while the eigenvector at the same time is an eigenvector with a pre-selectable quantum number s of a second operator S. Among the possible uses are the calculation of specific eigenvalues/ -vectors, convergence acceleration, coping with degeneracy or remedies to "pseudo convergence" [6].

If we think for example of H as being a Hamiltonian from the general structure of the problem and from the physical background convergence should be possible in a wide range. Since this note's purpose is to introduce and explain the algorithm, no formal proofs have been presented. Theoretical existence and convergence proofs may be based to a large extent on already existing theoretical knowledge especially about the generalized eigenvalue problem.

As a first "proof of concept" sample calculations were performed . It was possible to fully verify the expected behavior.

These encouraging results suggest to test the method in realistic cases. Since the algorithm is basically iterative it might as well be worthwhile to incorporate iterative methods for the solution of the inhomogeneous system of linear equations (cf. eq. (7b)) explicitly into the formulation to obtain new iterative prescriptions.

## Acknowledgement

The author thanks K. G. Kreuzer and H. G. Miller for helpful remarks and hints. The calculations were done using the package "FreeMath" available under the GNU public license.

# References

[1]  D K Faddejew and W N Faddejewa, Numerische Methoden der linearen Algebra (Oldenburg, München, 1964)
     A S Householder, The Theory of Matrices in Numerical Analysis (Blaisdell, New York, 1964)
[2]  Abstract Space: K G Kreuzer, H G Miller and W A Berger, Phys. Lett. A **81**, 429 (1981)
     Particle-hole representation/Nuclear Physics: R R Whitehead, Nucl. Phys. A **182** 290
     Particle-hole representation/Atomic Physics: W A Berger, PhD Dissertation, University of Frankfurt, 1979
     Coordinate Space: R C Andrew, H G Miller, Phys. Lett. A 318 , 487 (2003)
[3]  W A Berger, K G Kreuzer and H G Miller, Z. Phys. A 298, 11-12 (1980)
[4]  V Vorobyev, Method of Moments in Applied Mathematics ( Gordon and Breach, New York, 1962)
[5]  W A Berger, H G Miller K G Kreuzer and R M Dreizler, J. Phys. A **10**, 1089 (1977)
[6]  H G Miller and W A Berger, J. Phys. A **12**, 1693 (1979)
     W A Berger and H G Miller, J. Phys. A **40** 5675 ( 2007)